\documentclass[10pt,
reprint,
twocolumn,
amsmath,
amssymb,
aps,
prb,
superscriptaddress,
longbibliography]{revtex4-2}

\usepackage{graphicx}
\usepackage[T1]{fontenc}
\usepackage{dcolumn}
\usepackage{bm}
\usepackage{fancyhdr}

\usepackage[colorlinks,allcolors=blue]{hyperref}
\usepackage{xcolor}
\usepackage[utf8]{inputenc}

\usepackage{import}
\newcommand{\jh}[1]{\textcolor{black}{#1}}

\usepackage{hyperref}
\usepackage{textcomp}
\usepackage{siunitx}
\usepackage{upgreek}
\usepackage{transparent}
\usepackage{multirow}

\DeclareSIUnit\angstrom{\text{Å}}

\begin{document}

\preprint{APS/123-QED}



\title{Sub-10 nm Quantification of Spin and Orbital Magnetic Moment Across the Metamagnetic Phase Transition in FeRh Using EMCD}

\author{Jan Hajduček}
\thanks{Authors contributed equally.}
\affiliation{CEITEC BUT, Brno University of Technology, Purkyňova 123, 612 00 Brno, Czech Republic}

\author{Veronica Leccese}%
\thanks{Authors contributed equally.}
\affiliation{Institute of Physics (IPHYS), Laboratory for Ultrafast Microscopy and Electron Scattering (LUMES),
École Polytechnique Fédérale de Lausanne (EPFL), Lausanne 1015 CH, Switzerland}

\author{Ján Rusz}%
\affiliation{Department of Physics and Astronomy, Uppsala University, Box 516, 75120 Uppsala, Sweden}

\author{Jon Ander Arregi}%
\affiliation{CEITEC BUT, Brno University of Technology, Purkyňova 123, 612 00 Brno, Czech Republic}

\author{Alexey Sapozhnik}%
\affiliation{Institute of Physics (IPHYS), Laboratory for Ultrafast Microscopy and Electron Scattering (LUMES),
École Polytechnique Fédérale de Lausanne (EPFL), Lausanne 1015 CH, Switzerland}

\author{Jáchym Štindl}%
\affiliation{CEITEC BUT, Brno University of Technology, Purkyňova 123, 612 00 Brno, Czech Republic}

\author{Francesco Barantani}
\affiliation{Institute of Physics (IPHYS), Laboratory for Ultrafast Microscopy and Electron Scattering (LUMES),
École Polytechnique Fédérale de Lausanne (EPFL), Lausanne 1015 CH, Switzerland}

\author{Paolo Cattaneo}
\affiliation{Institute of Physics (IPHYS), Laboratory for Ultrafast Microscopy and Electron Scattering (LUMES),
École Polytechnique Fédérale de Lausanne (EPFL), Lausanne 1015 CH, Switzerland}

\author{Antoine Andrieux}
\affiliation{Institute of Physics (IPHYS), Laboratory for Ultrafast Microscopy and Electron Scattering (LUMES),
École Polytechnique Fédérale de Lausanne (EPFL), Lausanne 1015 CH, Switzerland}

\author{Vojtěch Uhlíř}
\affiliation{CEITEC BUT, Brno University of Technology, Purkyňova 123, 612 00 Brno, Czech Republic}

\author{Fabrizio Carbone}
\affiliation{Institute of Physics (IPHYS), Laboratory for Ultrafast Microscopy and Electron Scattering (LUMES),
École Polytechnique Fédérale de Lausanne (EPFL), Lausanne 1015 CH, Switzerland}

\author{Thomas LaGrange}
\email{thomas.lagrange@epfl.ch}

\affiliation{Institute of Physics (IPHYS), Laboratory for Ultrafast Microscopy and Electron Scattering (LUMES),
École Polytechnique Fédérale de Lausanne (EPFL), Lausanne 1015 CH, Switzerland}


\date{\today}

\begin{abstract}

Electron magnetic circular dichroism (EMCD) in transmission electron microscopy (TEM) enables element-specific measurement of spin and orbital magnetic moments, analogous to X-ray magnetic circular dichroism (XMCD). While the EMCD technique offers unmatched spatial resolution, its quantitative accuracy remains under scrutiny, particularly in beam-splitter geometries with convergent probes. Here, we systematically evaluate the limits of quantitative EMCD analysis using the first-order magnetostructural transition in the functional phase-change material FeRh as a tunable magnetic reference. Unlike previous EMCD studies primarily focused on elemental ferromagnets such as Fe, we demonstrate its applicability to a correlated material exhibiting coupled structural and magnetic order. We demonstrate that the extracted orbital-to-spin moment ratio ($m_\text{L}/m_\text{S}$) remains consistent with XMCD benchmarks for TEM probes down to approximately 6 nm, thereby establishing the validity range for reliable quantification. For nm-sized probes with higher convergence angles, we observe an enhanced $m_\text{L}/m_\text{S}$, which we attribute to a combination of instrumental factors and sensitivity to nanoscale heterogeneity within the probed volume. Our results confirm that EMCD provides quantitative agreement with macroscale techniques under suitable conditions, while uniquely enabling spatially confined measurements of local magnetic moments in functional magnetic materials, and allowing the study of interfacial, defect-mediated, or phase-separated magnetism that is inaccessible to photon-based methods.

\end{abstract}

\maketitle

\section{\label{sec:Intro} Introduction}

Understanding collective magnetic ordering at the nanoscale requires probes that combine spatial and elemental selectivity with sensitivity to local spin and orbital moments. Core-level excitations are especially powerful in this regard, as they directly connect microscopic electronic structure to emergent macroscopic magnetic behavior. Dichroic spectroscopies have therefore become indispensable tools for quantifying magnetic moments and exploring their spatial variation in correlated and functional materials.

\jh{X-ray magnetic circular dichroism (XMCD) is the established standard for element-specific magnetometry, in which polarization-dependent X-ray absorption spectra analyzed through the sum rules \cite{Thole1992, Carra1993} provide quantitative access to spin and orbital magnetic moments. Synchrotron-based imaging techniques extend XMCD to the nanoscale, with scanning transmission X-ray microscopy (STXM) achieving the visualization of magnetic nanoparticles down to approximately 15 nm in size \cite{Rosner2020SoftResolution}. Spatially resolved quantitative analysis based on XMCD sum rules has been demonstrated with lateral resolutions approaching 25 nm \cite{Robertson2015QuantitativeSpectro-microscopy}. Further improvement of spatial resolution in photon-based methods is fundamentally constrained by diffraction and optical focusing limits.}

Electron magnetic circular dichroism (EMCD) \jh{\cite{Hebert2003, Schattschneider2006}}, the electron-based analogue of XMCD, circumvents these spatial limitations by exploiting the advanced optics of modern transmission electron microscopes. Initially realized using a crystalline beam splitter \cite{Hebert2003,Schattschneider2006,Schattschneider2012} (for more details and the comparison with XMCD, see Appendix~\ref{Ap:A-EMCD}), EMCD has since diversified to convergent-beam geometries \jh{\cite{Schattschneider2008,Thersleff2017,Loffler2018Convergent-beamApplications}}, electron vortex beams \cite{Verbeeck2010, Uchida2010GenerationMomentum, McMorran2011ElectronMomentum}, orbital-angular-momentum sorters \cite{Rotunno2019}, and atomic-resolution \jh{scanning transmission electron microscopy - energy electron loss spectroscopy (STEM-EELS)} approaches \cite{Rusz2016MagneticResolution, Ali2020, Wang2018AtomicMicroscopy}. These methods allow quantitative access to spin and orbital moments through adapted sum-rule analyses \cite{Rusz2007, Calmels2007}, though experimental complexity and interpretation challenges persist, especially under low-symmetry or low-signal conditions \cite{Hebert2008MagneticSpectrometry,Fu2019ElectronStructure,Zeng2023TheDichroism}. While much of EMCD development has emphasized methodological advances, only a few studies have yet leveraged it to probe functional materials, such as \jh{in the case of the magnetic phase transition in the manganite} Pr$_{0.5}$Sr$_{0.5}$MnO$_3$ \cite{Rubino2008Energy-lossMicroscope}.

\begin{figure}[t]
\includegraphics[width = 1\columnwidth]{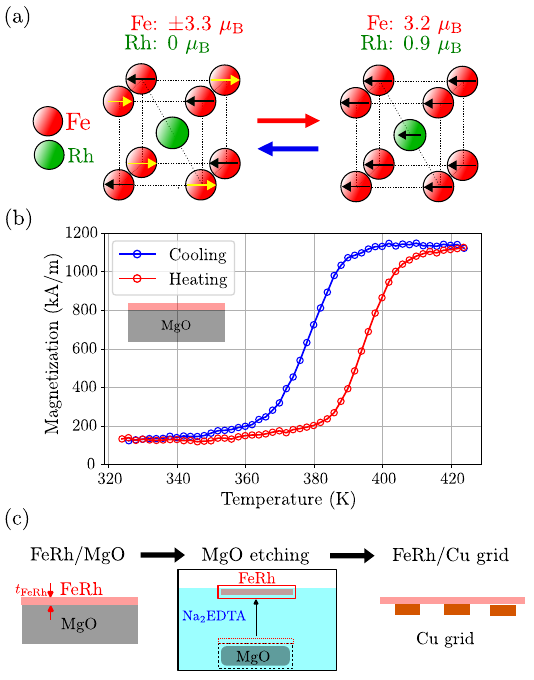}
\caption{\label{fig1} \textbf{FeRh system and studied specimen geometry} (a) Schematic of the AF-FM transition in FeRh, showing atomic positions and magnetic moments on Fe and Rh atoms. (b) \jh{Temperature-dependent magnetization during heating and cooling for the 25-nm-film on MgO before detachment.} (c) Sample preparation: epitaxial FeRh grown on MgO, followed by selective MgO etching in Na$_2$EDTA to release the FeRh film, which is then collected on a Cu TEM grid.} 
\end{figure}

\begin{figure} [t]
\includegraphics[width = 1\columnwidth]{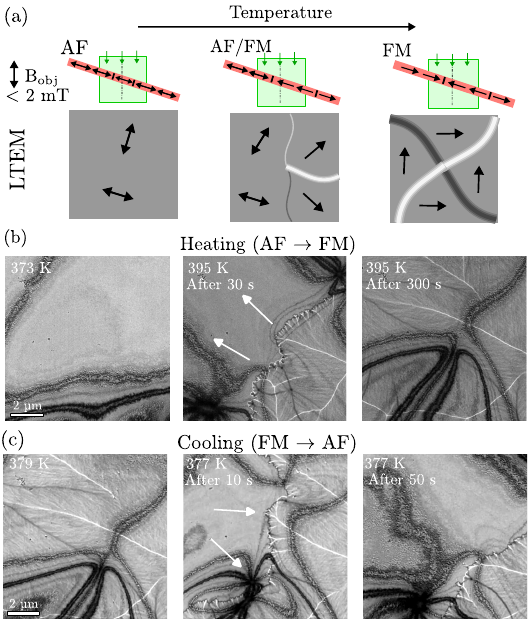}
\caption{\label{fig2} \textbf{AF-FM phase transition in LTEM.} (a) Schematic illustration of the pretilted freestanding FeRh (red box) illuminated by electron beam, with expected Fresnel contrast at AF–FM phase boundaries and FM domain walls. (b) In situ LTEM images showing the nucleation and growth of the FM phase during heating. (c) Corresponding retraction of the FM phase upon cooling, revealing the spatial evolution of the magnetic first-order phase transition.}
\end{figure}

\begin{figure*} [t]
\includegraphics[width = 2\columnwidth]{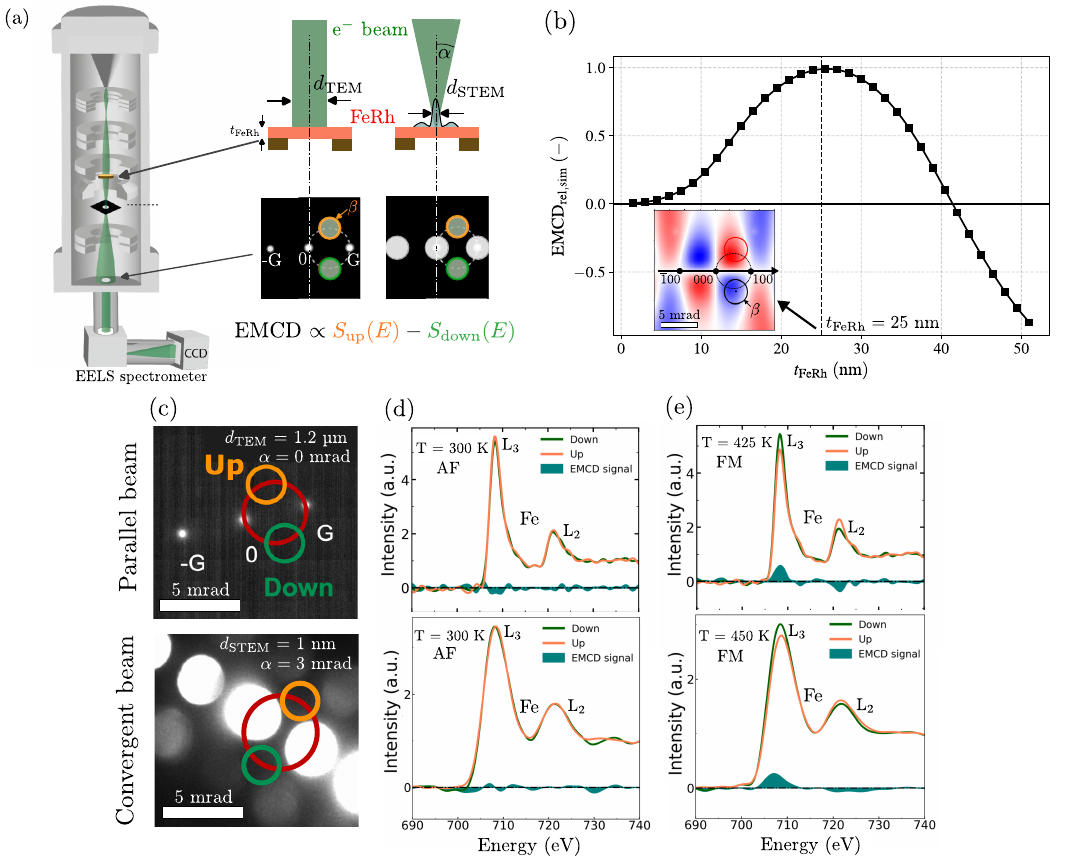}
\caption{\label{fig3} \textbf{EMCD in freestanding FeRh.} (a) Schematic of EMCD detection using parallel beam and convergent beam configurations. (b) Simulated relative EMCD signal as a function of sample thickness for a conventional experimental setup, showing optimal signal near 25~nm at 300~keV. (c) Experimental diffraction patterns with marked aperture positions used for dichroic signal acquisition on 25-nm-thick freestanding FeRh films. (d) EMCD spectra measured in the AF phase (300~K) for both configurations, showing weak \jh{noise-like differences}. (e) EMCD spectra in the FM phase (425–450~K), showing pronounced dichroic contrast.}
\end{figure*}

\jh{Here, we employ in situ EMCD to study the equiatomic FeRh alloy}, which undergoes a first-order \jh{phase} transition from antiferromagnetic (AF) to ferromagnetic (FM) order near 360 K \cite{Fallot1939, Maat2005}, with Fe moments of 3.3~$\mu_\mathrm{B}$ and negligible Rh moments in the AF phase, and Fe moments of 3.2~$\mu_\mathrm{B}$ and induced Rh moments of 0.9~$\mu_\mathrm{B}$ in the FM phase \cite{Shirane1964MagneticAlloys, Kunitomi1971DiffuseFeRh}, as indicated in Figure \ref{fig1}(a). \jh{The transition features interconnected structural, electronic, and magnetic degrees of freedom, and can be tuned by magnetic field, strain, spatial confinement, electrical currents, or optical excitation} \cite{Ohtani1994FeaturesFilm, Suzuki2009StabilityFilms, Lewis2016, Uhlir2016ColossalStripes, Arregi2020, Hajducek2025Dislocation-DrivenTransition}. Such properties make FeRh an interesting platform for nanoscale magnetic analysis \jh{across phase boundaries}.

We experimentally demonstrate sub-10 nm spatially resolved EMCD measurements in freestanding FeRh films across the metamagnetic phase transition upon in situ specimen heating. The dichroic signal is unambiguously assigned to intrinsic magnetic ordering, and sum-rule analysis yields $m_{\text{L}}/m_{\text{S}}$ ratios which are consistent with values obtained from XMCD experiments \cite{Stamm2008Antiferromagnetic-ferromagneticDichroism} for electron probes as small as 6 nm. We find that further probe miniaturization alters the apparent $m_{\text{L}}/m_{\text{S}}$ ratio, an effect we evaluate from both physical and instrumental perspectives. These results underpin the feasibility of nanoscale magnetic imaging with EMCD in functional materials. Combined with recent demonstrations of $m_{\text{L}}/m_{\text{S}}$ quantification at atomic resolution using large-convergence STEM-EMCD geometries \cite{Ali2025VisualizingMicroscope}, our study expands the scope of EMCD to serve as a tool for quantitative nanoscale analysis of fundamental magnetic properties in functional magnetic materials. Together, these advances establish EMCD as a robust probe of local spin and orbital magnetism, while delineating the experimental conditions under which quantitative analysis remains reliable and accurate.

\section{\label{sec:Results} Results}

\jh{The studied 25-nm-thick FeRh(001) films were epitaxially grown} on MgO(001) substrates by magnetron sputtering and subsequently prepared for TEM analysis by selective chemical etching (see experimental details in Appendix~\ref{Ap:B-Method}). The temperature-dependent \jh{magnetization data of the FeRh film on MgO}, measured by vibrating sample magnetometry (VSM), is shown in Figure \ref{fig1}(b), exhibiting thermal hysteresis of approximately 20~K and a residual magnetization in the AF phase of 150~kA/m. This confirms the high quality of the phase transition in deposited FeRh films used for the film detachment. Immersion \jh{in a 0.3 M solution of the disodium salt of ethylenediaminetetraacetic acid (Na2-EDTA) dissolves the MgO substrate \cite{Edler2010}} while preserving the integrity of the FeRh layer \cite{Motyckova2023}, which is then collected onto TEM-compatible Cu grids \cite{Hajducek2025Dislocation-DrivenTransition}, as schematically shown in Figure~\ref{fig1}(c).  

\begin{figure}
\includegraphics[width = 1\columnwidth]{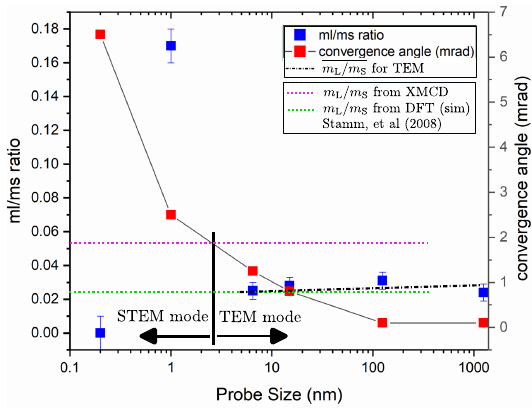}
\caption{\label{fig4}\textbf{Probe-size dependence of Fe magnetic moments in FeRh from EMCD.} Evaluated orbital-to-spin moment ratio ($m_\text{L}/m_\text{S}$) as a function of electron probe size. The red curve indicates the corresponding convergence angle of the beam. EMCD results are compared with reference XMCD values from Stamm \textit{et al.} \cite{Stamm2008Antiferromagnetic-ferromagneticDichroism}, showing good agreement for probe diameters larger than $\sim$6~nm in TEM mode. For smaller, highly convergent STEM probes, $m_\text{L}/m_\text{S}$ changes sharply, indicating limitations of quantitative extraction.} 
\end{figure}

The magnetic phase transition in the freestanding FeRh films was \jh{tracked} by in situ Lorentz transmission electron microscopy (LTEM) during the heating and cooling cycles. The field-free LTEM configuration enables direct visualization of FM domain formation without the need for external magnetic perturbation. The expected Fresnel contrast at AF–FM phase boundaries and FM domain walls is illustrated schematically in Figure~\ref{fig2}(a). As shown in Figure~\ref{fig2}(b), FM domains nucleate upon heating around 395~K and expand to form a single dominant phase domain, indicating a low density of nucleation centers within the film. In addition to the magnetic contrast, a pronounced evolution of diffraction contrast with temperature is observed, arising from strain induced by the lattice parameter change across the FeRh phase transition, which locally modifies the diffraction condition. The cooling series, as shown in Figure~\ref{fig2}(c), reveals the retracting FM contrast at approximately 377~K, \jh{with the film displaying a local phase transition thermal hysteresis of about 18 K, which is consistent with the magnetometry measurements} on the as-deposited film. \jh{This hysteretic behavior} confirms the first-order nature of the AF–FM transition in the freestanding films and reveals the stable nature of the magnetic phase evolution at the nanoscale.

To evaluate the spatial resolution and quantitative limits of EMCD, we performed probe-size-dependent measurements using the beam-splitter geometry, as schematically shown in Figure~\ref{fig3}(a) for parallel and convergent electron beams, along with the corresponding detection geometry in the diffraction plane. The detection is performed with an applied objective magnetic field of nearly 2 T, saturating the sample magnetization out of plane of the film. The crystal is also tilted to the 3-beam condition, ensuring the detectability of EMCD. The sample thickness represents a critical factor for EMCD detection, and its effect was evaluated based on the dynamical diffraction simulations, using the code from Ref. \cite{Rusz2017} with the simulation details summarized in \jh{Appendix~\ref{Ap:C-EMCD-sim}} and the result plotted in Figure~\ref{fig3}(b). The FeRh thickness dependence of EMCD is plotted by comparing the EMCD signals from two apertures in the diffraction plane constructed using the conventional \jh{Thales} circle geometry in a 3-beam configuration for a 300 keV parallel electron beam. A film thickness of approximately 25 nm provides the highest signals for reliable EMCD detection. The inset shows the corresponding EMCD signal distribution for 25 nm thickness. The accurate thickness of the prepared specimen \jh{was determined by X-ray reflectivity measurements of the FeRh film before detachment,} yielding \jh{$24.70\pm0.03$ nm} with an evaluated FeRh roughness of \jh{$0.13\pm0.02$ nm}. Figure \ref{fig3}(c) shows the experimental electron diffraction patterns with indicated collection apertures for EMCD detection for a 1.2-\textmu m-sized probe in TEM mode and for a 1-nm-sized probe in STEM mode. Aperture positions for dichroic signal collection were selected symmetrically with respect to the 100 systematic row \jh{axis}.

Experimental EMCD spectra from individual apertures and the differential spectrum at the Fe $L_{2,3}$ edges are presented in Figures~\ref{fig3}(d) and (e). The background of the collected spectra was subtracted, and the resulting spectra were normalized with respect to the post-edge region. The zero-loss peak was deconvoluted and smoothed using a Savitzky-Golay filter. In the AF phase (300~K) ofFeRh, the resulting dichroic signal is weak and approaches the noise floor, consistent with the negligible net magnetization. Upon heating the FeRh film into the FM phase (425–450~K), a pronounced EMCD contrast appears, exhibiting an apparent asymmetry at the L$_{3}$ and L$_{2}$ edges, enabling the quantitative extraction of spin and orbital contributions via sum-rule analysis.

The orbital-to-spin moment ratio can be evaluated using EMCD sum rules as \cite{Rusz2007, Calmels2007}


\begin{equation}
\frac{m_\text{L}}{m_\text{S}} = \frac{2 \, (p + q)}{3 \, (p - 2q)},
\label{eq1}
\end{equation}

\noindent where \(p=\int_{L_3} \Delta I(E) \, \mathrm{d}E\), and \(q=\int_{L_2} \Delta I(E) \, \mathrm{d}E\), with \(\Delta I(E) = I_{\text{Up}}(E) - I_{\text{Down}}(E)\) being the dichroic EMCD signal obtained by subtracting spectra measured at two symmetric aperture positions with respect to a mirror axis in reciprocal space.

The results of this analysis applied to our data are presented in Figure~\ref{fig4}, which shows the probe-size dependence of the extracted $m_\text{L}/m_\text{S}$ ratio, along with the corresponding convergence angle of the electron beam. For probe diameters as small as $\sim$6~nm, obtained by data acquisition in TEM mode, EMCD yields \(m_\text{L}/m_\text{S} = 0.026\pm0.005\), which remains consistent all the way up to the 1.2 \textmu m-sized probe. Experimental and simulated data for \(m_\text{L}/m_\text{S} \) in FeRh from Stamm \textit{et al.} \cite{Stamm2008Antiferromagnetic-ferromagneticDichroism}, also visualized in Figure~\ref{fig4}, show a systematic offset from our evaluated values. In contrast, for highly convergent STEM probes with diameters of \jh{sub-6~nm}, the ratio increases sharply to $0.17\pm0.01$. \jh{We observed that exposure under sub-nm focused beams induced sample damage}, which ultimately prevented us from evaluating the EMCD signal, for details see Appendix~\ref{Ap:D-Probe-damage}.

To contextualize these findings, Table~\ref{tab:table1} summarizes representative $m_\text{L}/m_\text{S}$ values and total magnetic moments \jh{obtained for the BCC Fe and FeRh systems via EMCD and XMCD experiments, along with the employed probe sizes.} BCC Fe serves as a reference system, showing systematically higher values of the measured $m_\text{L}/m_\text{S}$ value upon using EMCD as compared to XMCD. Considerable variations were observed for $m_\text{L}/m_\text{S}$ evaluated in atomically resolved EMCD \cite{Ali2025VisualizingMicroscope}, where the $m_\text{L}/m_\text{S}$ value averaged over multiple atomic planes was increased up to 0.16. On the contrary, we obtained systematically lower evaluated value of $m_\text{L}/m_\text{S}$ compared to corresponding XMCD value in FeRh reported in \cite{Stamm2008Antiferromagnetic-ferromagneticDichroism}. The potential origin of the observed variations of $m_\text{L}/m_\text{S}$ on Fe in FeRh is evaluated in the Discussion section from the perspectives of material properties and instrumental detection.

\section{\label{sec:Discussion} Discussion}
In this work we have demonstrated quantitative EMCD measurements of the AF–FM transition in freestanding FeRh films, enabling a direct comparison with macroscale XMCD results. On the 6 nm to 1~\textmu m scale, the extracted $m_\text{L}/m_\text{S}$ ratio stays within the range of $0.026\pm0.005$. These measurements reveal a systematic offset with respect to the corresponding experimental XMCD \cite{Stamm2008Antiferromagnetic-ferromagneticDichroism}, while matching well with the theoretical value of 0.023 obtained from density functional theory calculations \cite{Stamm2008Antiferromagnetic-ferromagneticDichroism}. When employing highly convergent STEM probes, substantial deviations in $m_\text{L}/m_\text{S}$ are observed. These differences can arise from both instrumental effects related to beam confinement and data acquisition, as well as from intrinsic nanoscale variations in the material.

\begin{table}[t]
\caption{Comparison of selected literature and current work values for orbital-to-spin moment ratio \( m_{\text{L}}/m_{\text{S}} \), total magnetic moment \( m_{\text{tot}} \), and probe size for Fe and FeRh systems using XMCD and EMCD. TW stands for "This work".}
\centering
\begin{tabular}{|c|c|c|c|c|}
\hline
\multirow{2}{*}{Material} & \multirow{2}{*}{Technique} & $m_{\text{L}}/m_{\text{S}}$ & Probe size & \multirow{2}{*}{Ref.} \\
 & & $[-]$ & [\textmu m] & \\
\hline
\hline
\multirow{13}{*}{\textbf{BCC Fe}} & \multirow{4}{*}{XMCD} & 0.043 & $> 10^{2}$ & \cite{Chen1995ExperimentalCobalt} \\
 & & 0.03 & $>10^{3}$ & \cite{Honghong2005X-raySubstrate} \\
 & & 0.07 & $-$ & \cite{OBrien1994ExperimentalMagnetism} \\
 & & $0.12 \pm 0.05$ & $-$ & \cite{Zaharko2001SoftTransmission} \\
\cline{2-5}
 & \multirow{9}{*}{EMCD}& $0.09 \pm 0.03$ & $-$ & \cite{Calmels2007} \\ 
&& $0.08 \pm 0.01$ & $\sim10^{-3}$ & \cite{Lidbaum2009QuantitativeMicroscopy} \\
&& $0.065 \pm 0.005$ & $-$ & \cite{Warot-Fonrose2010EffectFilm} \\ 
&& $0.04 \pm 0.01$ & $-$ & \cite{Rusz2011InfluenceMeasurements} \\ 
&& 0.08 & $\sim10^{-3}$ & \cite{Thersleff2015QuantitativeDichroism} \\
&& \jh{$0.06\pm 0.03$} & $\sim5\cdot10^{-4}$ & \cite{Rusz2016} \\
&& $0.06\pm 0.01$ & $\sim$ & \cite{Ali2019QuantitativeEELS} \\
&& $0.08-0.09$ & $< 5\cdot10^{-3}$ & \cite{Ali2021} \\
&& 0.16 \jh{\cite{Ali2025-comment}} & $\sim10^{-4}$ & \cite{Ali2025VisualizingMicroscope} \\
\hline
\hline
\multirow{3}{*}{\textbf{FeRh}} & XMCD & 0.054 & $>10^{2}$ & \cite{Stamm2008Antiferromagnetic-ferromagneticDichroism} \\
\cline{2-5}
 & \multirow{2}{*}{EMCD} & $0.026\pm 0.005$ & 1.2 & TW \\
 & & $0.17 \pm 0.01$ & $1\cdot10^{-3}$ & TW \\
\hline
\end{tabular}

\label{tab:table1}
\end{table}

On the instrumental side, several factors are known to influence EMCD quantification. The preferential three-beam geometry provides optimal symmetry of the EMCD signal, but asymmetries between the signal lobes below and above the systematic row are known to occur and may bias the extracted $m_\text{L}/m_\text{S}$ values \cite{Lidbaum2009QuantitativeMicroscopy}. Additionally, the convergence angle plays a critical role. Even though the low convergence angles, below 1 mrad, preserve the conventional collection geometry and minimize distortions \cite{Thersleff2015QuantitativeDichroism}, it is known that overlapping diffraction disks lead to less straightforward interpretation \jh{\cite{Loffler2018Convergent-beamApplications, Rusz2017LocalizationResolution}}. Even at higher convergence angles, quantification is still possible but requires careful treatment of geometries deviating from the standard Thales circle condition \cite{Rusz2016, Ali2025VisualizingMicroscope}. Additional sources of error include plural scattering in EELS, which artificially increases $m_\text{L}/m_\text{S}$ if the low-loss spectrum is untreated \cite{Rusz2011InfluenceMeasurements, Zeng2023TheDichroism}, as well as the finite energy resolution of the spectrometer and aberrations in STEM optics. 

Beyond instrumental factors, intrinsic material properties play a crucial role in confined-probe EMCD measurements. When the probe size falls below $\sim$10~nm, it interrogates regions comparable in scale to individual dislocations or phase boundaries, where local strain and magnetization variations can modify the dichroic response (see Appendix~\ref{Ap:E-Disl}). At ultrasmall probe sizes, additional effects such as beam-induced damage and local changes in orbital magnetism become significant, as observed in our atomically resolved analysis (see Appendix~\ref{Ap:D-Probe-damage}). Localized strain gradients from the defect network can alter the dynamical diffraction condition and geometries essential for measuring EMCD signals with the beam splitter technique. For measurements with small probe sizes, we intentionally positioned the probe away from areas of strong diffraction contrast that could perturb the measurement. Such limitations may be mitigated by using an atomic-scale electron vortex probe EMCD \jh{\cite{Verbeeck2010, Rusz2014a, Pohl2017AtomMomentum}}.    

Furthermore, reduced dimensionality—such as surfaces and interfaces—has been shown to enhance orbital contributions. Atomically resolved EMCD in Fe has revealed site-dependent variations in $m_\text{L}/m_\text{S}$ and orbital moment enhancement at surfaces \cite{Ali2025VisualizingMicroscope}, consistent with long-standing observations of increased orbital magnetism in reduced-dimensional systems \cite{Gambardella2003GiantNanoparticles, Tischer1995EnhancementCu100, Xu2001GiantFe/GaAs100, Autes2006MagnetismWire, Edmonds2001SizeParticles, Desjonqueres2007OrbitalDimensionality}. Additional influences, including partial film oxidation \cite{Thersleff2015QuantitativeDichroism} or local magnetic anisotropies, cannot be excluded.

\jh{An open question} remains on how far the usual approximations for extracting the orbital-to-spin moment ratio $m_\text{L}/m_\text{S}$ can be pushed across systems ranging from atomically localized electrons to fully itinerant Bloch states. Many materials, featuring mixed valence, charge- or spin-density waves, and nanoscale multiferroicity, blur this boundary. In principle, the XMCD and EMCD sum rules are valid for both localized and itinerant electrons because they relate integrated dichroic intensities to ground-state expectation values for transitions between well-defined shells (e.g., 2p$\rightarrow$3d in transition metals) and are largely model-independent \cite{Irkhin2005SumFerromagnets, vanderLaan2014X-rayMagnetism}. In practice, sum rules developed and benchmarked on local-moment systems are often applied to metallic solids with care as well. Microscopically, the notion of a "local moment" and its link to valence is subtle: high-energy, ultrafast probes such as XMCD/EMCD effectively take snapshots of the system in the presence of a deep core hole \cite{Braicovich2005FemtosecondEmission, Braicovich2008RationaleExcitation}, which can transiently suppress valence fluctuations characteristic of metallic bonding and thus reveal the instantaneous local moment on the ion. For quantitative work, potential corrections (e.g., $j$-$j$ mixing, the magnetic dipole term, and—in EMCD—dynamical diffraction) should be assessed. A potential future direction for EMCD research and theory is to push toward true atomic-scale resolution, enabling site-resolved measurements of $m_\text{L}/m_\text{S}$ across unit cells \cite{Ali2025VisualizingMicroscope}, \jh{impurities \cite{Boske1994CircularNi}}, defects, and interfaces to disentangle localized versus itinerant contributions and, in turn, empirically test and refine the limits of the sum-rule approximations in complex materials. 

Taken together, these results establish EMCD as a~robust, element-specific probe of magnetic moments. Quantitative reliability is demonstrated for parallel and moderately confined probes, while increased sensitivity at the nanometer scale reveals subtle heterogeneities in functional magnetic materials. By systematically identifying the experimental regime in which EMCD results coincide with photon-based XMCD, this study outlines the conditions for accurate quantification. It highlights the unique capability of EMCD to probe magnetism at spatial resolutions far exceeding those of photon-based methods.

\section{\label{sec:Conclusion} Conclusion}

In this work, we have demonstrated that EMCD enables quantitative analysis of the AF–FM transition in freestanding FeRh with nanometer spatial resolution. The results were directly benchmarked against XMCD. For TEM probes down to $\sim$6~nm, the extracted $m_\text{L}/m_\text{S}$ ratio remains consistent with established values obtained by XMCD, confirming the validity range of EMCD for quantitative studies. At higher convergence angles, deviations emerge, which we attribute to a combination of instrumental effects and intrinsic sensitivity to local variations within the probed region.

These findings highlight that EMCD not only provides quantitative agreement with bulk-sensitive XMCD under suitable conditions but also reveals enhanced sensitivity to nanoscale heterogeneity such as strain, defects, and surface contributions. This capability is consistent with atomically resolved EMCD studies \cite{Ali2025VisualizingMicroscope} and establishes EMCD as a powerful, element-specific probe of confined or inhomogeneous magnetic phases. By leveraging standard TEM infrastructure, EMCD bridges the gap between bulk-averaged spectroscopies and nanoscale magnetism, opening new opportunities for studying functional magnetic materials.

\begin{acknowledgments}
 This work was supported by the project TERAFIT No. CZ.02.01.01/00/22$\_$008/0004594. Access to the CEITEC Nano Research Infrastructure was supported by the Ministry of Education, Youth and Sports (MEYS) of the Czech Republic under the project CzechNanoLab (LM2023051). J.H. was supported by the Thermo Fisher Scientific scholarship. V.L., A.S., F.B., P.C., A.A., F.C., and T.L. acknowledge the support by the Air Force Office of Scientific Research under award number FA8655-24-1-7015. J.R. acknowledges the support of the Swedish Research Council (grant no.\ 2021-03848), the Olle Engkvist's Foundation (grant no.\ 214-0331), and the Knut and Alice Wallenberg Foundation (grant no.\ 2022.0079). The simulations were enabled by resources provided by the National Academic Infrastructure for Supercomputing in Sweden (NAISS) at the NSC Centre, partially funded by the Swedish Research Council through grant agreement no.\ 2022-06725.

\end{acknowledgments}

\section*{\label{sec:DATA} Authors contribution statement}

V.L. and T.L. performed the EMCD measurements and electron microscopy characterization. V.L., A.S., and T.L. analyzed the data. J.H. and J.R. performed the EMCD simulations. J.A.A. and V.U. provided the FeRh materials and characterization. J.H. drafted the paper. T.L. and F.C. supervised work. All authors contributed to the interpretation of the data and the refinement of the paper. 

\section*{\label{sec:DATA} DATA STATEMENT}

All data and code used to generate the presented figures are available in the Zenodo repository \cite{DOI:10.5281/zenodo.17301295}.



\appendix

\begin{figure*}
\includegraphics[width = 1.7\columnwidth]{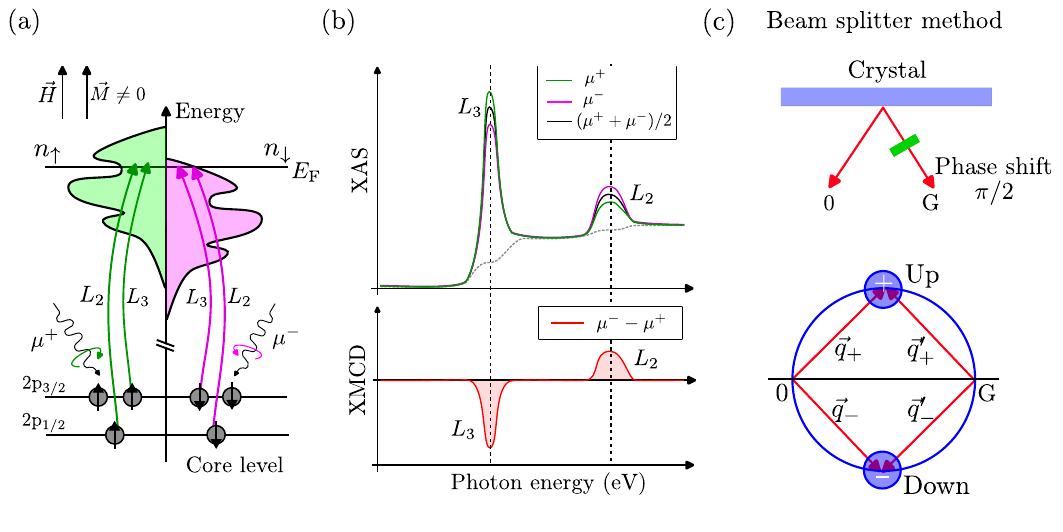}
\caption{\label{fig:EMCD-XMCD} \textbf{Interlink between XMCD and the beam splitter EMCD method.} (a) Schematic of core level spin-orbit coupled electron transitions \jh{excited by} different handedness circular polarized X-rays. (b) Example X-ray absorption spectra (XAS) and difference (XMCD) spectrum for Co. Adapted from \cite{vanderLaan2014X-rayMagnetism}. (c) Schematic of the beam splitter method for EMCD and Thales circle diagram. Adapted from \cite{Schattschneider2006}.} 
\end{figure*}

\section{\label{Ap:A-EMCD} EMCD BEAM SPLITTER METHOD}

EMCD is the electron-scattering analogue of XMCD, both probing element-specific magnetic properties through spin–orbit–coupled electronic transitions. In XMCD, dichroism arises from the differential absorption of left- and right-circularly polarized X-rays whose photon helicity couples to the sample magnetization (see Figure~\ref{fig:EMCD-XMCD}(a) and (b)). In conventional EMCD, an equivalent helicity is generated by the phase difference between symmetrically scattered electron beams under controlled diffraction conditions in a TEM (Figure~\ref{fig:EMCD-XMCD}(c)). This interference produces an effective circular polarization of the electron’s momentum transfer, leading to an energy-loss asymmetry analogous to the XMCD effect. Both techniques obey the same dipole selection rules ($\Delta m=\pm 1$) and share a common theoretical foundation based on angular momentum transfer between the probe and the magnetized electrons. While XMCD detects real-photon absorption in the electromagnetic regime, EMCD measures virtual-photon exchange through inelastic electron scattering, offering identical magnetic information with nanometer to atomic spatial resolution. 

In this work, we employed a method commonly referred to in the literature as the intrinsic method, or beam splitter method \cite{Schattschneider2006}. In general, for performing these experiments, the following four conditions must be met:

\begin{enumerate}
    \item \jh{An EMCD experiment requires at least two electron plane-wave components to interact simultaneously with the magnetic sample.}
    \item The two momentum transfers from the two electron beams must not be parallel to measure EMCD. Ideally, the two momentum transfers should be perpendicular to each other.
    \item \jh{The two momentum transfers must not be in-phase and, ideally, have a phase shift of $\pm\pi/2$ (the optimal condition for measuring the magnetic circular dichroism).}
    \item \jh{To measure the \jh{EMCD spectral} difference in the scattering, we must be able to change the helicity of the excitation.}
\end{enumerate}

This approach involves using a crystalline sample to split the electron beam via Bragg scattering, fulfilling conditions 1 and 3 simultaneously. Bragg scattering divides the incident electron beam into two beams, which also induces a phase shift that can be optimized ($\pi$/2) by controlling the diffraction conditions and sample thickness. The method can satisfy conditions 2 and 4 by performing momentum-resolved electron energy loss spectroscopy ($q$-EELS) and by selecting appropriate scattering directions of left-hand and right-handed polarization directions, e.g., selecting those defined by the Thales circle in the diffraction plane in main text Figure \ref{fig3} with the spectrometer entrance aperture.

\begin{figure*}[t]
\includegraphics[width = 1.8\columnwidth]{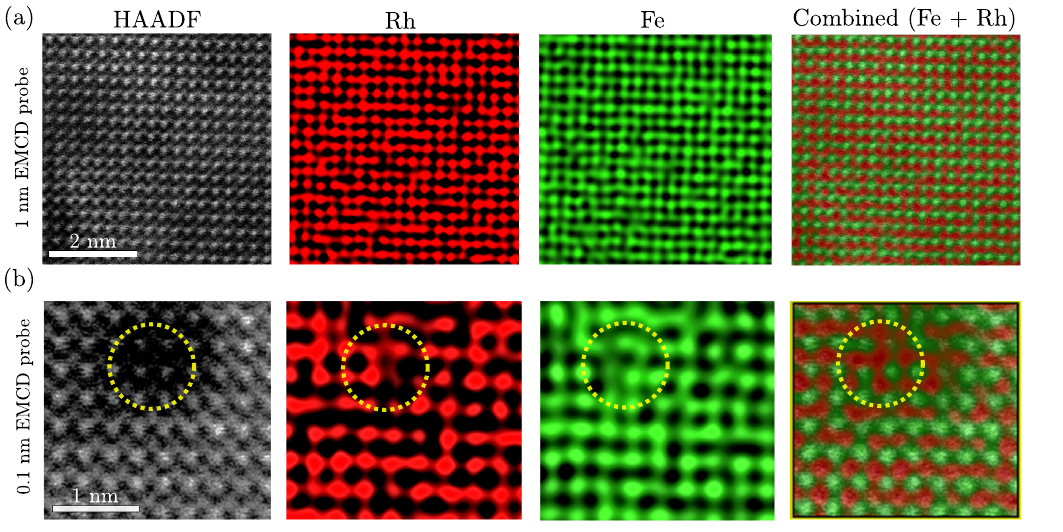}
\caption{\label{fig:S1} \textbf{Beam damage induced by small EMCD probes.} Atomically resolved HAADF and EDS maps of FeRh after 15 min exposure to focused probes of different diameters: (a) 1 nm and (b) 0.1 nm, with the beam-exposed areas marked by yellow circles. Individual EDS maps for Rh (red) and Fe (green), together with the combined Fe + Rh distribution, are shown alongside the corresponding HAADF images. The 1 nm probe leaves no detectable footprint, whereas the 0.1 nm probe produces a distinct dark spot in the HAADF image and a locally damaged Fe–Rh structure, indicating the onset of beam-induced degradation under sub-nanometer probe confinement.} 
\end{figure*}

\section{\label{Ap:B-Method} EXPERIMENTAL METHODS}

\jh{FeRh films of 25 nm thickness were deposited from an~equiatomic target on MgO(001) substrates using a~magnetron sputtering system (BESTEC) with a~base pressure lower than 5$\times$10$^{-8}$ mbar. The substrates were preheated to 723~K for 1~h before deposition and film growth was carried out at the same temperature at an Ar pressure of 2.7$\times$10$^{-3}$ mbar and a~deposition rate of 0.3 \si{\angstrom}/s. The films were post-growth annealed at 1000 K for 30 min to obtain the desired CsCl-type crystallographic structure. The samples were then cooled in high vacuum and capped with a~3-nm-thick Al layer at room temperature. The thickness of the films was determined by X-ray reflectivity (Rigaku SmartLab 9 kW diffractometer with Cu K$_{\alpha}$ radiation, $\lambda$ = 1.5406 \si{\angstrom}). Temperature-dependent magnetization data was acquired via vibrating sample magnetometry (Quantum Design VersaLab) in an applied magnetic field of 3 T. The field-induced shift of the phase transition temperature (corresponding to $-8$~K/T) was compensated for in the dataset, and magnetization data are presented after subtracting the diamagnetic substrate contribution.}

Lorentz TEM was performed on a Titan Themis at 300~kV using a Gatan heating holder in field-free mode (<1 mT, with the objective lens off). Fresnel contrast was acquired at 1 mm defocus during heating and cooling cycles with a stability of $\pm 0.5$ K. The AF–FM transition was reproducibly cycled more than 20 times, and the transition temperature was determined with an accuracy of $\pm 1$ K, with an additional calibration uncertainty of $\pm 5$ K relative to the VSM. Magnetic fields up to $\pm1.8$~T could be applied along the beam axis by exciting the objective lens.

\begin{figure*}[t]
\includegraphics[width = 1.8\columnwidth]{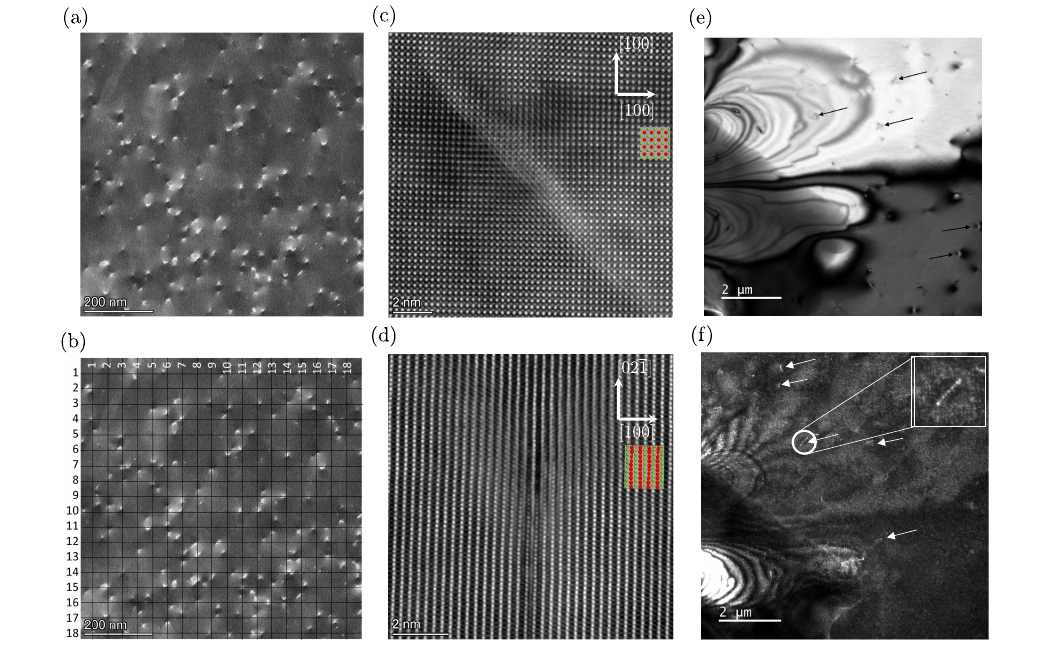}
\caption{\label{fig:FeRh-defects} \textbf{\jh{Dislocation density analysis in FeRh thin films}.} (a) Off-zone axis HAADF-STEM image showing a uniform dislocation density of $\sim 5 \times 10^{9}$ cm$^{-2}$. (b) Same image with line-intercept grid used for quantitative density evaluation. (c)~High-resolution HAADF-STEM image of dislocations from panel (b) at higher magnification. (d) HAADF-STEM image along the [012] zone axis showing an a[100] edge dislocation; insets in (c,d) show colored Fe (green) and Rh (red) lattice overlays on [001] and [012] projections, respectively. (e,f) TEM bright-field and dark-field images under two-beam conditions highlighting individual dislocations (arrows). The inset in \jh{(f)} shows a magnified dislocation with dotted-line contrast, characteristic of a~screw-type defect.} 
\end{figure*}

EMCD spectra were recorded in the same instrument equipped with an X-FEG and GIF spectrometer. Data were acquired in scanning mode to generate $30 \times 30$ spectral cubes with a total integration time of approximately $\sim$15 min. Probe sizes of 1.2 \textmu m (TEM mode) and 1 nm (STEM mode) were employed. In STEM, drift correction ensured sub-nm stability, while in TEM the drift remained below 5 nm. A dispersion of 0.25 eV/channel over 2048 channels (vertically binned) yielded an effective energy resolution of 1–3 eV, broadened by spectrometer aberrations in STEM. Background subtraction was performed using a double arc tangent function fit, and spectra were normalized after the edge. Integration windows for the Fe $L_{2,3}$ edges were set to 705–711 eV and 718–724 eV in TEM, and 704–714 eV and 716–726 eV in STEM to \jh{account for peak broadening, that occurs as a consequence of leaving the optimized condenser–objective alignment regime where the STEM column is aberration-corrected to the decreased convergence angle}.

High-resolution structural TEM analysis was performed using the same instrument with double aberration correction.

\section{\label{Ap:C-EMCD-sim} EMCD SIMULATIONS}

Dynamical diffraction simulations of the EMCD signal were performed using a parallel electron beam, with the \jh{\textsc{mats}} code first presented in \jh{\cite{Rusz2013}} with the improved summation regime from \jh{\textsc{mats.v2}} method~\cite{Rusz2017}.

\jh{Thickness dependent} diffraction patterns were calculated on a reciprocal grid spanning \jh{$\pm10$~mrad} in both $q_x,q_y$ with a \jh{0.25}~mrad step \jh{and with 1.49 nm thickness step}. The incident beam \jh{had the energy of} 300~keV ($\lambda = 1.968$~pm) \jh{and passed through the FeRh crystal oriented in a three-beam orientation with Bragg spots \textbf{G}=$\pm(100)$, obtained by tilting approximately 9.5° from the zone axis (001). The beam is therefore aligned along the (016) zone axis direction of the crystal. The Bloch wave coefficient excitation threshold was set to $10^{-4}$. Resulting momentum-resolved EMCD magnetic signal was normalized by a corresponding non-magnetic signal.}

\section{\label{Ap:D-Probe-damage} BEAM DAMAGE UPON USING SMALL EMCD PROBES}

To assess the effect of probe confinement on specimen stability, we compared the structural response of freestanding FeRh films to prolonged illumination by focused electron probes. Atomically resolved STEM experiments were performed using probe sizes of 1~nm and 0.1~nm, with identical exposure times of 15~minutes at 300~kV. High-angle annular dark-field (HAADF) imaging and energy-dispersive X-ray spectroscopy (EDS) elemental mapping were acquired after exposure to evaluate possible beam-induced damage. It is also important to note that both the electron dose and flux is two orders of magnitude higher for the 0.1 nm probe, which is expected to produce significant sputtering of the film and disorder.

As shown in Figure \ref{fig:S1}(a), the 1~nm probe does not produce any observable footprint in either HAADF contrast or Fe/Rh elemental distributions, confirming that EMCD experiments with this probe size are non-destructive under the applied conditions. In contrast, Figure \ref{fig:S1}(b) shows the corresponding analysis for the 0.1~nm EMCD probe, which results in a visible dark spot in HAADF images, together with a distorted atomic arrangement and altered Fe–Rh distribution in EDS maps. These observations demonstrate that sub-nanometer probes can cause irreversible structural modifications, while nanometer-scale probes provide a safe regime for quantitative EMCD measurements in FeRh.

\section{\label{Ap:E-Disl} Dislocation density analysis}

The density and character of dislocations in the freestanding FeRh films were investigated by combining TEM and STEM imaging. Conventional TEM bright-field and dark-field images recorded under two-beam diffraction conditions along the $\langle 110 \rangle$ direction enhance defect contrast, allowing identification of isolated dislocations (Figure~\ref{fig:FeRh-defects}(e,f)). The inset in Figure~\ref{fig:FeRh-defects}\jh{(f)} shows a~magnified view of a dislocation with dotted-line contrast, consistent with a screw-type dislocation. Due to dynamical scattering effects and limited sample tilting, not all dislocations are visible under a~given two-beam condition, and bend contours or secondary phase contrast may obscure weaker features. As a~result, TEM-based counting tends to underestimate the dislocation density.

In comparison, STEM imaging with convergent-beam illumination provides enhanced and more isotropic dislocation visibility (Figure~\ref{fig:FeRh-defects}(a–d)). In this mode, the averaging of intensities over a range of incidence directions reduces strain-related background contrast, making individual dislocations more easily distinguishable. High-resolution HAADF-STEM further resolves the dislocation cores, with Figure~\ref{fig:FeRh-defects}(c,d) showing edge-type a[100] dislocations viewed along different zone axes, with corresponding colored Fe (green) and Rh (red) lattice overlays.

To quantify the dislocation density, we applied the line-intercept method \jh{\cite{Smith1953MeasurementSectioning,Ham1961TheFilms}}, which estimates the density $\Lambda$ as:
\begin{equation}
    \Lambda = \dfrac{2N}{tL},
\end{equation}
where $N$ is the number of intersections between dislocations and an overlaid grid (Figure~\ref{fig:FeRh-defects}(b)), $L$ is the total grid length, and $t$ is the local foil thickness. Applying the method to our STEM images yields an average dislocation density of $\sim 5 \times 10^{9}$ cm$^{-2}$ in the freestanding FeRh films.

\bibliography{references-main}

\end{document}